\documentclass[]{spie}  

\usepackage{amsmath,amsfonts,amssymb}
\usepackage{graphicx}
\usepackage[colorlinks=true, allcolors=blue]{hyperref}
\usepackage{caption}
\usepackage{subcaption}
\usepackage{rotating}

\title{Experimental validation of active control of low-order aberrations with a Zernike sensor through a Lyot coronagraph.}

\author[a]{Raphaël Pourcelot}
\author[a]{Mamadou N'Diaye}
\author[b]{Emiel H. Por}
\author[b]{Marshall Perrin}
\author[b]{Rémi Soummer}
\author[c,f]{Iva Laginja}
\author[b]{Ananya Sahoo}
\author[a]{Marcel Carbillet}
\author[b]{Greg Brady}
\author[b]{Matthew Maclay}
\author[b]{James Noss}
\author[b,d]{Pete Petrone}
\author[b]{Laurent Pueyo}
\author[b,g,h]{Scott D. Will}

\affil[a]{Université Côte d'Azur, Observatoire de la Côte d'Azur, CNRS, Laboratoire Lagrange, France}
\affil[b]{Space Telescope Science Institute, 3700 San Martin Drive, Baltimore, MD 21218, USA}
\affil[c]{Aix Marseille Univ., CNRS, CNES, LAM, Marseille, France}
\affil[d]{Hexagon Federal, Chantilly, VA 20151, USA}
\affil[f]{DOTA, ONERA, Université Paris Saclay, F-92322 Châtillon, France}
\affil[g]{Institute of Optics, University of Rochester, Rochester, NY 14620, USA}
\affil[h]{NASA Goddard Space Flight Center, Greenbelt, MD 20771, USA}

\newcommand{\up}[1]{\textsuperscript{#1}}

\begin{document} 
\maketitle

\begin{abstract}
Future large segmented space telescopes and their coronagraphic instruments are expected to provide the resolution and sensitivity to observe Earth-like planets with a 10\up{10} contrast ratio at less than 100 mas from their host star. Advanced coronagraphs and wavefront control methods will enable the generation of high-contrast dark holes in the image of an observed star. However, drifts in the optical path of the system will lead to pointing errors and other critical low-order aberrations that will prevent maintenance of this contrast. To measure and correct for these errors, we explore the use of a Zernike wavefront sensor (ZWFS) in the starlight rejected and filtered by the focal plane mask of a Lyot-type coronagraph. In our previous work, the analytical phase reconstruction formalism of the ZWFS was adapted for a filtered beam. We now explore strategies to actively compensate for these drifts in a segmented pupil setup on the High-contrast imager for Complex Aperture Telescopes (HiCAT). This contribution presents laboratory results from closed-loop compensation of bench internal turbulence as well as known introduced aberrations using phase conjugation and interaction matrix approaches. We also study the contrast recovery in the image plane dark hole when using a closed loop based on the ZWFS. 

\end{abstract}


\section{INTRODUCTION}
\label{sec:intro}  

Detecting extra-solar planets has become usual nowadays. Since the first detection in 1995 \cite{Mayor1995}, more than 4000 have now been identified. As the technique improves, one of the main goal is now to find Earth-like planets around Sun-like stars. In this investigation, collecting direct photons from the planet can prove to be extremely useful. Through spectroscopy they can unveil precious information on the planet atmospheric composition, effective temperature and surface gravity. However, in addition to be faint objects, up to 10\up{10} times fainter than their host stars, Earth twins are expected to appear at angular separations smaller than 100 milliarcseconds from their star. To provide this resolution and sensitivity, a new generation of large telescopes will be required. Such observatories are under study and could be part of the NASA's missions for the coming decades. For example  Habitable Exoplanet Imaging Mission (HabEx) \cite{Habex2019} or the Large Ultraviolet Optical Infrared surveyor (LUVOIR) \cite{LUVOIR2019} space observatories, among the proposals for the 2020 Astrophysics Decadal Survey, will propose telescopes which primary mirror diameter could reach respectively 4 m and 15 m. To envision such large telescopes, one option is to consider the segmentation of the primary mirror to be folded in a shuttle for launch and deployment in space. 

Nevertheless, the flux ratio between the host star and the planet, expected to reach up to 10\up{10}, still has to be overcome to identify the planet photons without fail. One way to remove stellar light is coronagraphy, a technique that removes most of the light from an observed star within the instrument using a succession of amplitude and/or phase masks. On top of these elements, deformable mirrors (DM) are used with stroke minimization algorithms \cite{Pueyo2009, Mazoyer2018_1, Mazoyer2018_2, Will2021} to shape the wavefront in order to create a dark area or Dark Hole (DH) in the star image to enhance the planet detection. If such coronagraphs can theoretically provide the required starlight extinction for exo-Earth detection \cite{Ruane2018,Guyon2003,N'Diaye2016,Seo2019}, their implementation needs an extreme optical stability thorough the observations. This stability decrease with the telescope size and design complexity (e.g. segmentation of the primary mirror). In the case of LUVOIR A, the 15-m version of the observatory, the wavefront error should not evolve of more than 10 pm RMS every 10 minutes \cite{Ultra2019}. Among the critical wavefront errors that need to be tightly constrained, we will focus on the low-order aberrations. They are mainly due to pointing or focusing errors, originating from thermal of mechanical drifts in the optical train. 

In the case of the LUVOIR A-like design, a Lyot Coronagraph (LC) is considered \cite{N'Diaye2016, Seo2019}. The LC uses amplitude masks among which the Focal Plane Mask (FPM), an opaque mask to suppress the on-axis star light. It can be implemented by using a mirror with a pinhole that suppresses starlight by letting it through while reflecting the off-axis planet light. Using a Zernike Wavefront Sensor (ZWFS) \cite{Wallace2011, N'Diaye2013} in this rejected and low-pass filtered beam, we explore strategies to retrieve the relevant information of the optical aberrations. This device has shown capabilities to measure aberrations with a precision down to the picometric level\cite{Guyon2005, Ruane2020}, making it an ideal candidate to measure the low wavefront error drifts. The ZWFS is currently considered for the measurement of low order aberrations in the context of the Roman Space Telescope (RST) Coronagraphic instrument (CGI) \cite{Shi2016}. 

To validate our concept studies, we use the facilities of the High-Contrast imager for Complex Aperture Telescopes (HiCAT) \cite{hicat1, hicat2, hicat3, hicat4, hicat5, hicat6} located at the Space Telescope Science Institute (STScI) in Baltimore. Rather than achieving the 10\up{10} raw contrast, this testbed aims to develop and validate the system approach (combination of coronagraphy, deformable mirrors and wavefront sensors for a segmented-like aperture telescope). After describing the ZWFS setup in the coronagraph rejected light on HiCAT, we will present some of the recent results, such as closed-loop correction of the bench internal turbulence with different strategies. Finally, we will show the correction of introduced Zernike aberrations, in term of wavefront error and contrast recovery in the final image plane .  

\section{LOWFS ON HICAT} 
\subsection{ZWFS as a low-order wavefront sensor}
To achieve their raw contrast goal over the exposure times, the next generation of large space telescopes will require a picometric stability \cite{Ultra2019}. Among the different regimes of aberrations, in term of both spatial and temporal frequencies, the low-order aberrations, such as line-of-sight or focusing errors and quasi-static aberrations have been identified as critical to maintain the contrast in the image DH. On top of research on the structures themselves, a key aspect is  wavefront control, or active optics. It uses several devices, such as DMs and precise WFSs to measure and correct for the wavefront error drifts in real time. With the LC, such as the Classical LC (CLC) or the Apodized Pupil LC (APLC), the FPM rejects the on-axis stellar light through its pinhole, yielding a high-flux beam that carries these low-order aberrations. On HiCAT, several strategies are under study to make the most of this beam: phase-retrieval, Point Spread Function (PSF) tracking with a Target Acquisition (TA) camera, a quad-cell sensor for fast and precise tip/tilt measurement and a ZWFS, that we are using in our experiments. For the experiments presented in this work HiCAT has been used in monochromatic light at 640 nm.    
\subsection{ZWFS formalism} \label{sec:formalism}

\begin{figure}
    \centering
    \includegraphics[width=.7\columnwidth]{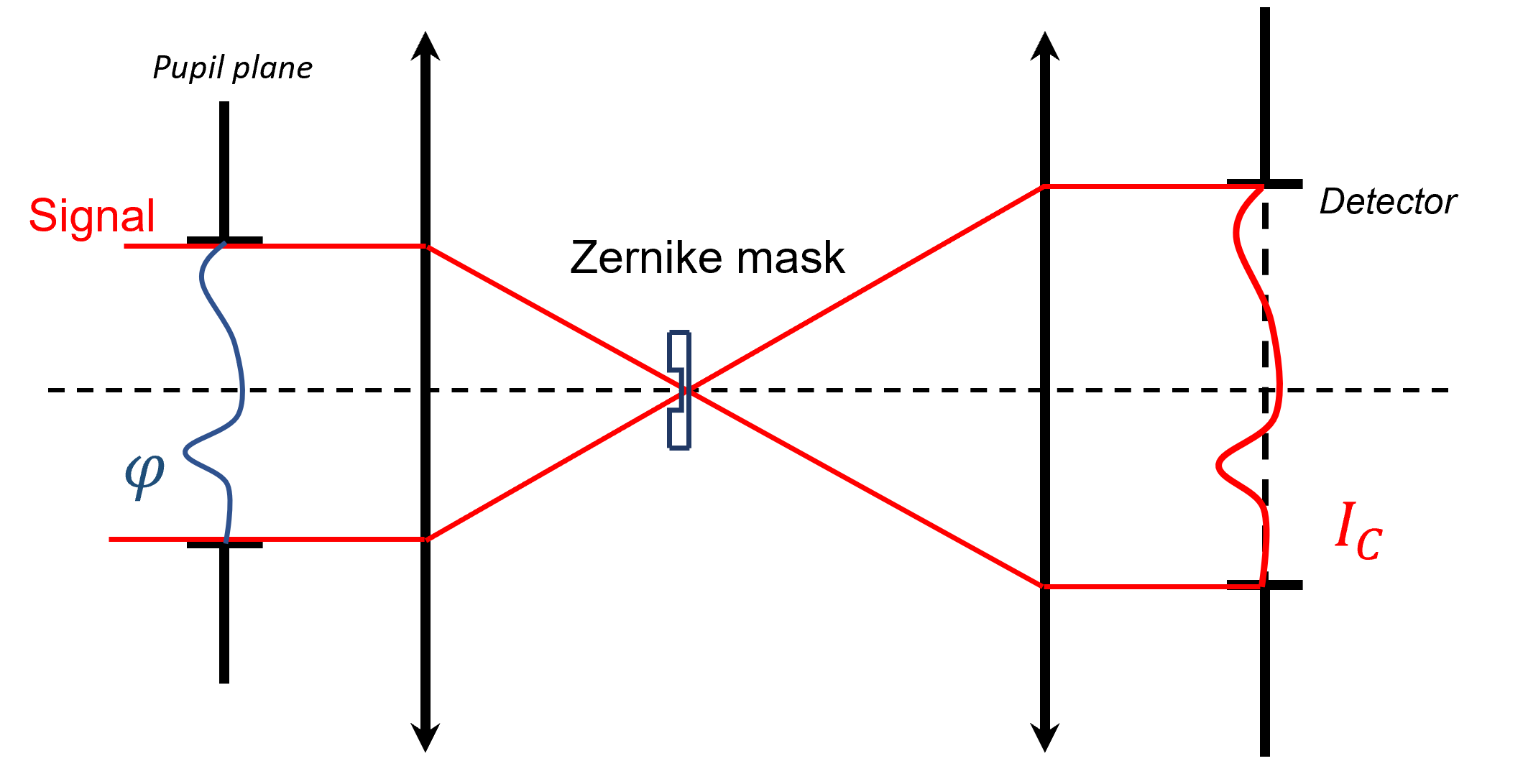}
    \caption{Scheme describing the ZWFS principle. }
    \label{fig:zwfs_scheme}
\end{figure}

Presented as the ultimate sensor in terms of sensitivity \cite{Guyon2005, Ruane2020, Chambouleyron2021}, the ZWFS acts as a common-path interferometer: a phase dimple on a silicate substrate in focal plane phase shifts part of the light, creating two superimposing beams, one delayed by a quarter-wave with respect to the other. These then interfere in the following pupil plane to encode the entrance phase errors as an intensity as described on Fig.~\ref{fig:zwfs_scheme}. We recall here the formalism described in several works \cite{Wallace2011, N'Diaye2013}.

In the general case, filtered or not, the intensity $I_C$ for each pixel on the pupil camera can be expressed as a function of the phase error $\varphi$:
\begin{align}
    I_C = \left[P\cos\varphi - b(1-\cos\theta)\right]^2 + \left[P\sin\varphi+b\sin\theta\right]^2,
\end{align}
\noindent
with P the pupil amplitude before the ZWFS, $\theta$ the phase shift introduced by the Zernike mask and b estimation of the pupil plane amplitude due to the light shifted by the dimple, and supposed constant through the acquisitions. 

Under the assumption of small phase errors $\varphi << 1$, this equation can be inverted to retrieve $\varphi$ as a function of $I_C$. While there exists an exact solution \cite{Ruane2020}, a simpler linear or second-order approximations have been derived in previous papers. In this paper we will use the latter that yields in the general case:

\begin{align}
    I_C = P^2 + 2b^2(1-\cos\theta)+2Pb\left[\varphi\sin\theta-(1-\varphi^2/2)(1-\cos\theta)\right] \label{eq:analytic_formula}.
\end{align}

In the HiCAT case where the beam is filtered this expression can be used to estimate the filtered wavefront error $\varphi_C$ \cite{Pourcelot2020}. It requires considering the filtered pupil $P_C$ instead of $P$. The expression of $P_C$ can be expressed as a convolution product of $P$ and the Fourier transform of transmission of the FPM denoted $\widehat{M_C}$:
\begin{align}
    P_C = P \otimes \widehat{M_C}.
\end{align}
The filtered phase $\varphi_C$ also writes as:
\begin{align}
    \varphi_C = \varphi \otimes \widehat{M_C}.
\end{align}

The reconstruction of the wavefront is given by the solution of Eq.~\ref{eq:analytic_formula} obtained through the computation of its discriminant $\Delta$:
\begin{align}
    \Delta = \sin^2{\theta} - \frac{2 \cdot (b-P_C)\cdot (1-\cos^2{\theta})}{P_C}  - \frac{P_C^2-I_E\cdot(1-\cos{\theta})}{P_C\cdot b} .
\end{align}

The solution then writes as
\begin{align}
    \varphi_C = \frac{-\sin{\theta}+\sqrt{\Delta}}{1-\cos{\theta}}.
\end{align}

\subsection{Hardware implementation and control algorithms} 

\begin{sidewaysfigure}
    \centering
    \centerline{\includegraphics[width=1\linewidth]{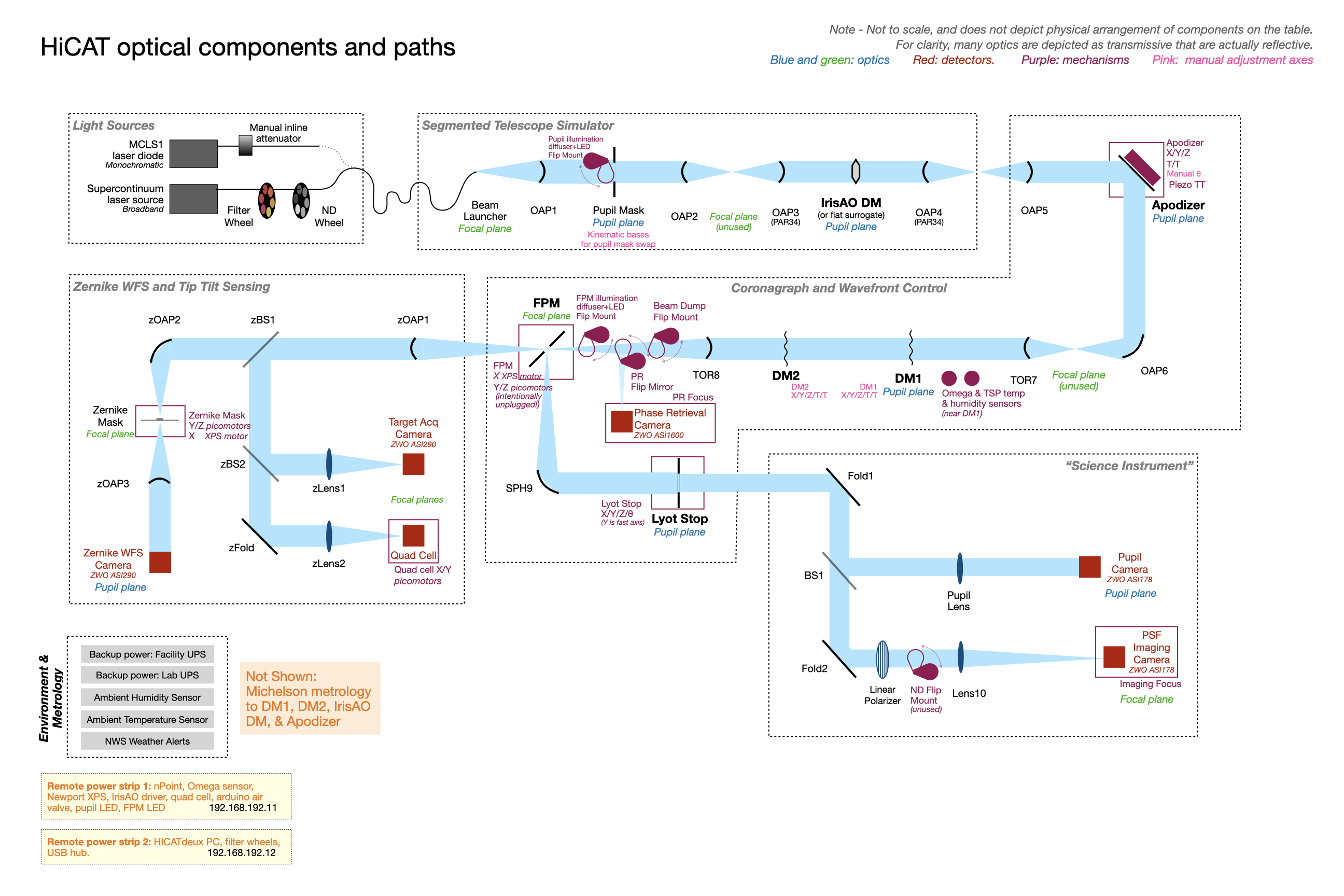}}
    \caption{Optical layout of the HiCAT testbed. The ZWFS is located on the left side with the Zernike mask and its dedicated camera.}
    \label{fig:blueprint}
\end{sidewaysfigure}

In the context of future large missions with exoplanet imaging capabilities, HiCAT aims at maturing coronagraphs for complex apertures telescopes. On top of a segmented deformable mirror to mimic segmented aperture telescopes, HiCAT has installed two continuous Boston kilo-DMs: DM1 in pupil plane and DM2 out of pupil plane for phase and amplitude control of the electric field. Designed to implement in the end an apodizer to turn it into an APLC, the bench currently uses a CLC for coronagraph: this includes a FPM and a Lyot stop \cite{Lyot1932}. We operate the ZWFS in the path where the source light is rejected by the coronagraph FPM. A blueprint of the HiCAT implementation is presented in Fig.~\ref{fig:blueprint}. The ZWFS is located on the left side, behind the FPM that acts as a spatial low-pass filter on the transmitted light at a cut-off frequency of 4.26 cycles/pupil. The ZWFS phase dimple has an angular diameter of $1.016\lambda/D_{pup}$, $\lambda$ being the working wavelength of 640 nm and $D_{pup}$ the entrance pupil diameter. It has a depth of 280 nm yielding a phase shift of $\lambda/5$ at 640 nm.

The control of the testbed as well as the analysis algorithm are using Python as a programming language. In particular, the hardware interface is based on the public package catkit \cite{catkit_0.36.1}, developed at STScI. For the experiment presented in this paper, we have been using an integrator controller to correct for the drifts with respect to a reference state measured at the beginning of each experiment. While we use DM2 to introduce knowned aberrations, only DM1 is currently used for correction with ZWFS control. The procedure is the following: at each iteration $k$, a measurement is performed by acquiring an image on the ZWFS camera. The reference measurement is subtracted, and the corresponding DM1 command update is added to the previous DM command with a gain $\gamma$. To get the best corrections possible, we investigate two different ways of measurements. The first strategy consists in performing phase conjugation: inverting Eq. \ref{eq:analytic_formula} allows for the reconstruction of a phase map. The DM command is therefore computed in two steps: by downsizing the phase map from the camera pixel size to match the DM actuator size; then by applying a numerical Fourier low-pass filter at the same cutoff frequency as the FPM's to prevent from introducing wavefront errors with unwanted spatial frequencies. The second strategy is to use an interaction matrix also called Jacobian matrix built on a predefined basis set. In this paper the convenient Zernike polynomial basis is used to poke DM1. Both strategies have advantages and drawbacks that will be detailed later.   

\subsection{Metrics and analysis} 

To assess the performance of our closed-loop correction, several metrics can be used. First we use the analytical Optical Path Difference (OPD) map computed with Eq.~\ref{eq:analytic_formula}. From this OPD, we can estimate the standard deviation of the wavefront error as well as an estimation of Zernike mode coefficients with an appropriated projection. Similarly, if an interaction matrix approach is used, the evolution of the coefficients sent to the DM give an overview of the correction quality. Since the ZWFS works in a differential way by subtracting the reference measurement to the actual measurements, wavefront errors specific to the ZWFS optical train should not disturb the measurements. To assess the relevance of the measurement and control, we have the ability to use other cameras for diagnostic. For example, the TA camera, a focal plane camera located in a beam close to the ZWFS allows for centroid tracking of the PSF. Using it, we can estimate the remaining tip/tilt aberration component of the wavefront error. Finally in the context of exoplanet imaging, an important metric is the focal plane contrast in the dark zone, that can be monitored. The DH images are presented in contrast scale. They are not representative of the current ultimate performance of HiCAT.

\section{CLOSED-LOOP ON BENCH INTERNAL TURBULENCE}
\begin{figure}
    \centering
    \includegraphics[width=\columnwidth]{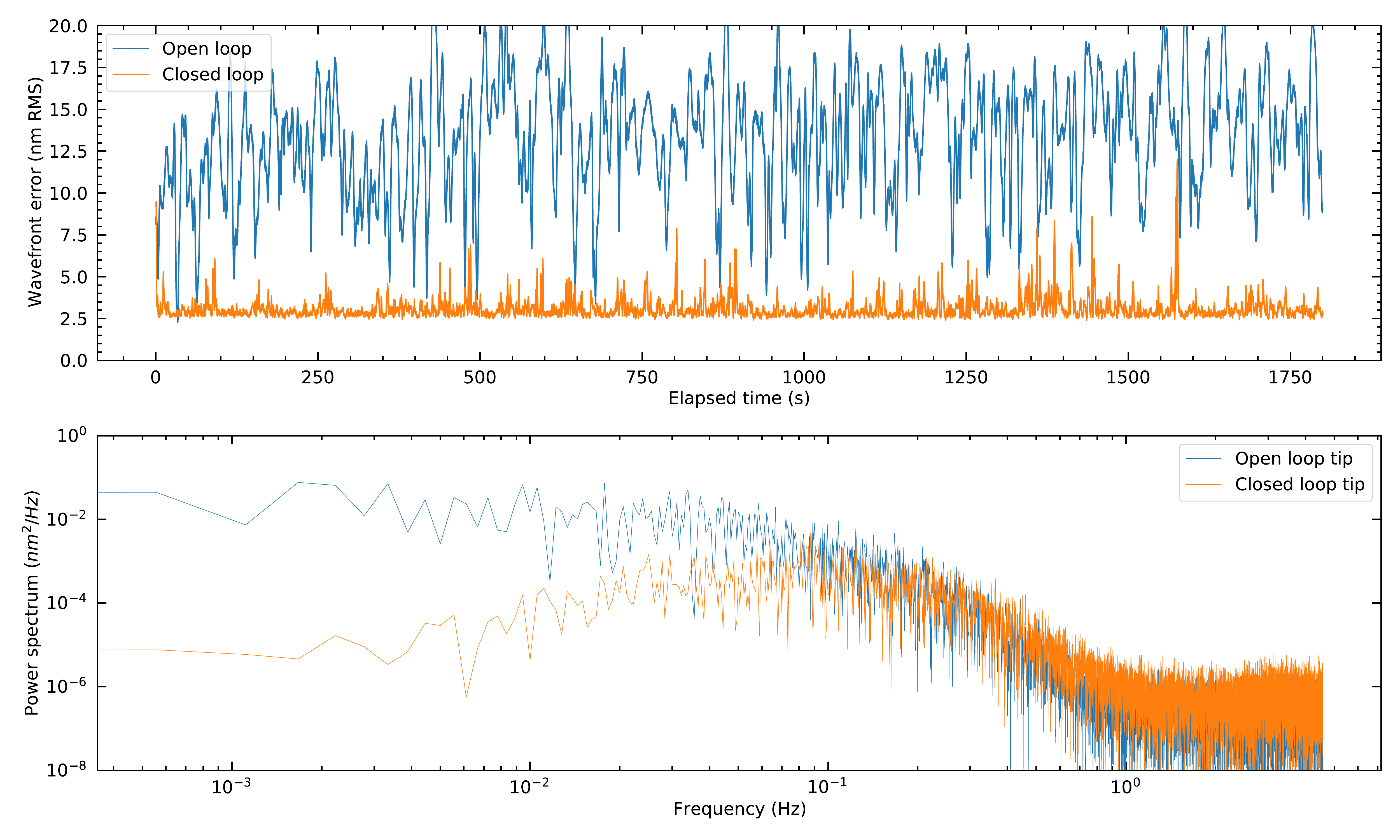}
    \caption{\textit{Top:} Evolution of the standard deviation computed by the ZWFS in an open loop experiment (in blue), and in a closed loop experiment with the Jacobian matrix approach (in orange). \textit{Bottom:} PSDs of tip mode. The control was performed with an interaction matrix built on 35 modes, 25 being kept in the inversion.}
    \label{fig:long_run}
\end{figure}

HiCAT is located in a clean room with temperature and humidity control, that allows for a stable environment. On top of that, an enclosure prevents the external turbulence from affecting the operations. Nevertheless, there are persisting air motion that translates into low-order aberrations. An overview of the turbulence as measured by the ZWFS is presented in Figure~\ref{fig:long_run}. It shows the measurement of the RMS wavefront error measured by the ZWFS, in open loop (in blue), as well as the Power Spectrum Density (PSD) in the bottom plot. The turbulence happens mainly in low frequencies: most of the turbulence is under 1 Hz. The typical amplitude of the drifts is of the order of magnitude of 10 nm RMS in the OPD. 

One of the experiments that can be run with the ZWFS on HiCAT is internal turbulence measurement and correction. Located behind the FPM, the ZWFS does see the turbulence integrated from source to FPM plus the turbulence between the FPM and the ZWFS. If this presents non-common path aberrations with the focal plane camera, it is still possible to investigate on ways to mitigate this turbulent motion. Depending on the light source setup, and therefore flux, it is possible to run the ZWFS camera at temporal frequencies ranging from 25 Hz to 60 Hz, allowing for a good sampling of the turbulence. 

\begin{figure}
    \centering
    \includegraphics[width=.5\columnwidth]{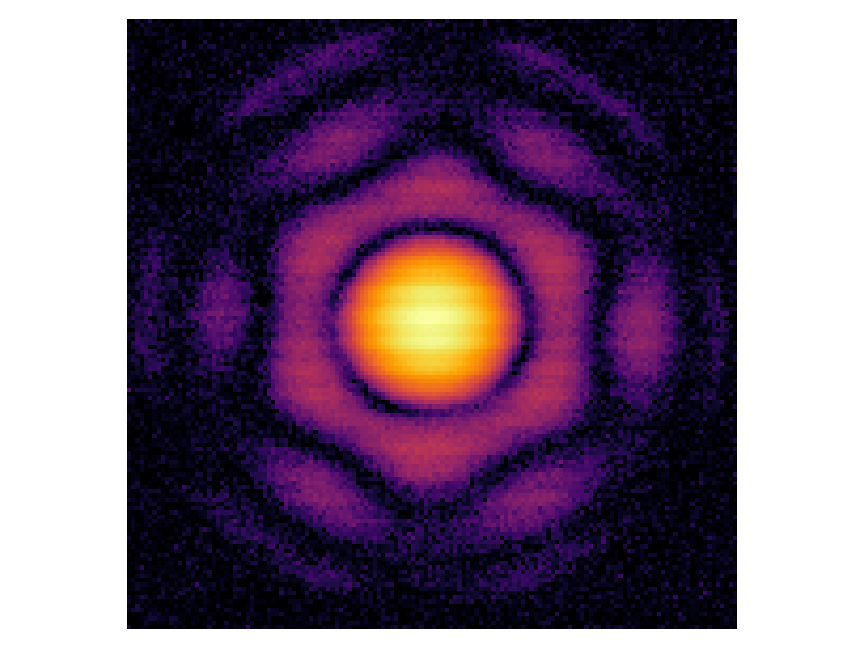}
    \caption{Example of TA image with vertical banding due camera readout, in logarithmic scale.}
    \label{fig:ta_image}
\end{figure}

\begin{figure}
    \centering
    \includegraphics[width=\columnwidth]{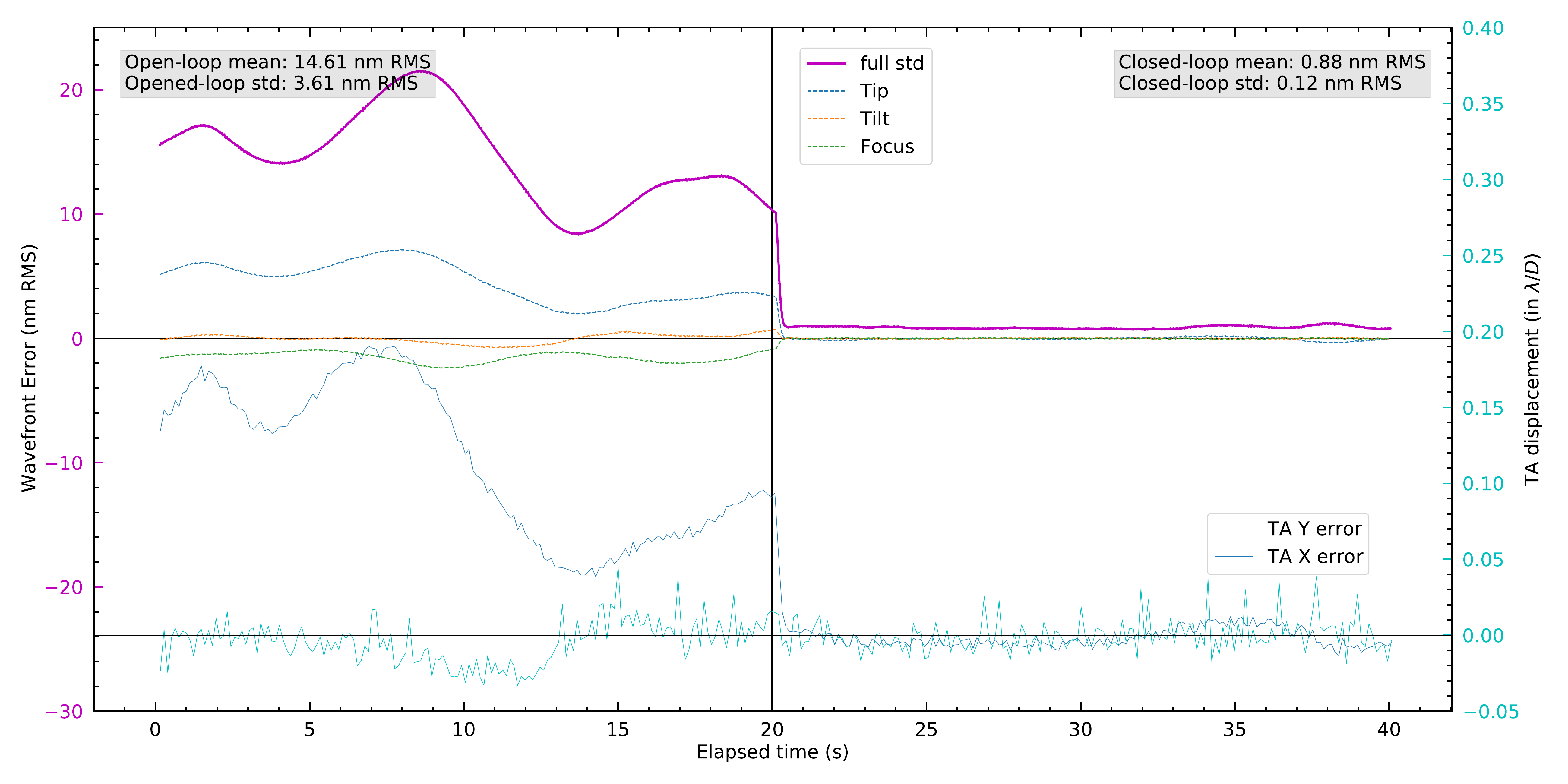}
    \caption{Evolution with time of the RMS wavefront error (purple) as well as the fitted tip (blue), tilt (yellow) and focus (green) coefficients in open loop (left part) and in close loop (right part), with a Jacobian matrix approach. The loop was running at 25 Hz. The blue and cyan curves show the position PSF centroid from the TA camera, and refer to the right vertical axis. }
    \label{fig:closed_loop}
\end{figure}

Closing the loop on turbulence shows promising results, as displayed on Fig.~\ref{fig:closed_loop}. On this plot, the control method used was a Jacobian built on 35 modes, 18 being kept in the inversion. The left section of the plot shows drifts in the low-order modes. The dominant coefficient is tip, that is associated with horizontal air motion on the bench. With the HiCAT enclosure we indeed expect stronger horizontal than vertical airflow. At first glance, PSF centroid position correlates well with the tip/tilt aberrations that is common-path turbulence located before the FPM. On the opposite, the PSF centroids show a jitter, especially in the vertical direction, that does not appear in the fitted Zernike coefficients (in blue, yellow and green). We attribute these to banding effect due to TA camera readout, creating vertical artifacts as visible on Fig.~\ref{fig:ta_image}. Some acoustic vibrations of elements on the bench at higher temporal frequencies have also been detected by other wavefront sensors. Finally, this might be coming from non-common path turbulence as well. The right part of Fig.~\ref{fig:closed_loop} shows the closed loop behavior. The low order coefficients in the OPD have there been accurately corrected. In this configuration, we reach an average mean value of 0.88 nm RMS, mostly dominated by fitting errors. The temporal standard deviation of the WFE shows promising results with a value of or 120 pm over 20 s. This correction is also stable on longer time periods, as Fig.~\ref{fig:long_run} depicts: the loop remains stable over 30 minutes. In this longer experiment the mean residual WFE at 3.1 nm RMS is dominated by fitting error, ie the aberrations that are not controlled by the command matrix. Despite this residual mean error, it presents a standard deviation of 640 pm RMS over the 1800 s. The closed loop PSD shows an improvement from one to four orders of magnitude between 0 and 0.1 Hz. This stability paves the way for future operations such as having the ZWFS control loop running as a background process while performing other bench operations.


\begin{figure}
    \centering
    \includegraphics[width=\columnwidth]{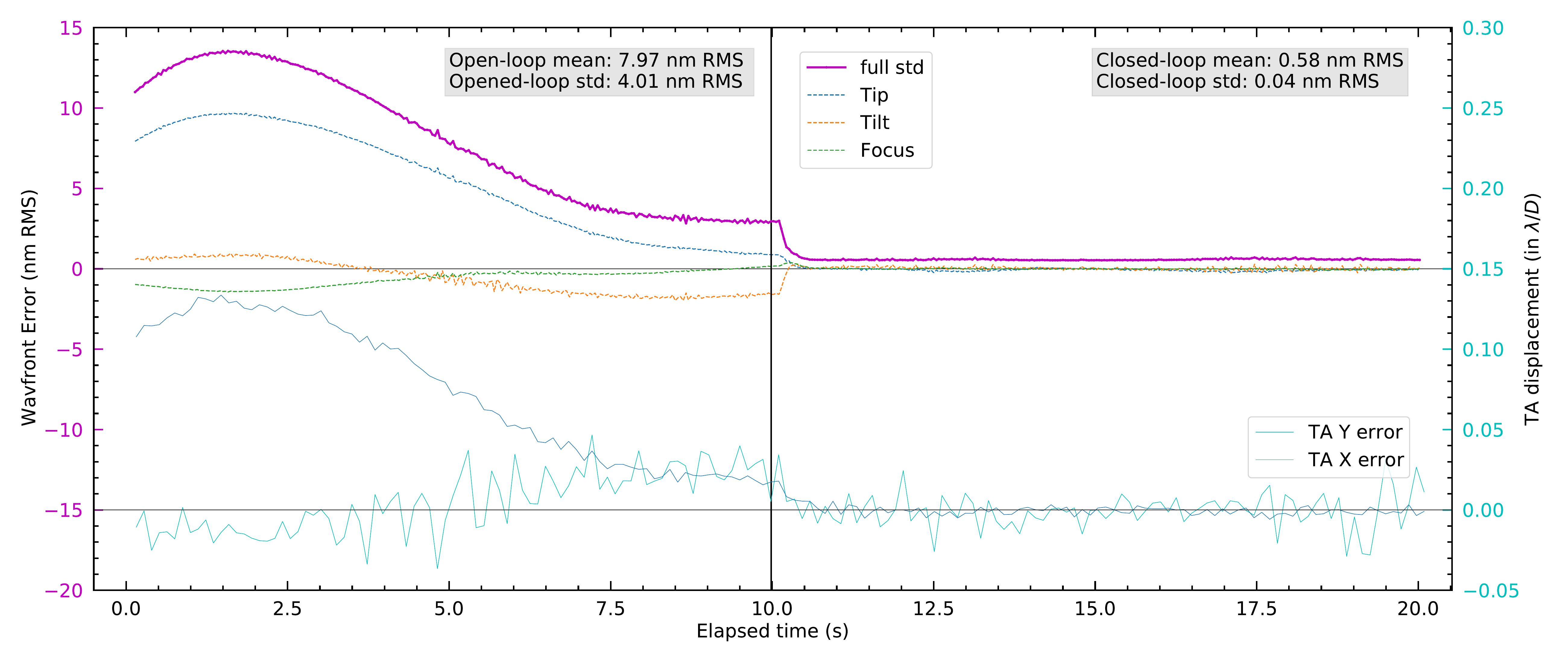}
    \caption{Same as Fig.~\ref{fig:closed_loop} but with a phase conjugation approach and a loop running at 25 Hz.}
    \label{fig:closed_loop_pc}
\end{figure}

It is interesting to note that the phase conjugation approach have very similar results as the Jacobian matrix approach. As shown in Fig.~\ref{fig:closed_loop_pc}, the same kind of stability can be reached. The phase conjugation shows a mean residual WFE at 0.58 nm RMS and a temporal standard deviation of the WFE on the order of 40 pm RMS over the 10 s of closed loop. The low-order coefficients fitted in the OPD are also well corrected. The residual error in closed loop comes here from calibration error. As detailed in Sec.~\ref{sec:formalism}, the analytical OPD computation requires the estimation of both the pupil amplitude $P$ and the reference wave $b$ amplitude, deduced from the calibration frame acquired at the beginning of the experiment. Errors due to air drifts during this calibration, or amplitude drifts during operation will translate into phase estimation error. This effect will also impact the Jacobian matrix approach, but it is currently dominated by fitting error. Further tests will be performed on the mitigation of these effects. Currently five references frames are averaged, but more could me used to increase the signal-to-noise ratio. It is important to note that the statistics are not strictly comparable, as the experiments were performed on different days with possibly different experiment conditions and the parameters such as the durations are not equal. Finally, WFE estimation is performed from the analytical OPD in both cases even if the control with the Jacobian matrix approach directly uses the camera images, introducing a possible bias in the convergence goal. Further experiments will be conducted to enable a fair comparison of the two approaches.

If interaction matrices have proven robust for many adaptive optics systems, some trade-offs still have to be found to achieve an optimal control of the low-order wavefront errors. While using a local interaction matrix where each of the 952 actuator of DM1 is poked would probe unseen spatial frequencies because of FPM filtering, the choice of using Zernike mode presents also drawbacks. Once filtered, this basis set is not a basis anymore of the pupil aberrations which might create redundancies in the probes aberration space. This could create the fitting error observed in Fig.~\ref{fig:closed_loop} if too few modes are used, typically less than 20. On the contrary, if too many modes are included, the input modes will also include wavefront errors with larger spatial frequencies unseen by the ZWFS that will be introduced in the system and not controlled. These will degrade the operation of the other elements of the bench and lead to loop divergence. Further investigations are therefore necessary.
As for phase conjugation, it presents a user-friendly method that does only require the sensor calibration frames. The long term stability of such a method will be explored in further studies.

\section{CLOSED-LOOP CONTROL ON INTRODUCED ABERRATIONS}

A key aspect of the ZWFS is to allow for stable DH images over time. At the current level of HiCAT operations, the raw contrast is typically maintained at $3\times 10^{-8}$ in the DH in monochromatic mode. With the CLC at these performance, the degradation of contrast due to air turbulence is not barely noticeable at timescale of seconds. To study the effect of ZWFS corrections on the focal plane DH, we introduce stronger aberrations with a deformable mirror. Using the ZWFS measurement, we then try to recover the initial contrast in the focal plane image. In this experiment, the aberration introduced on DM1 was 25 nm RMS of astigmatism. To avoid introducing spatial frequency components that cannot be seen by the ZWFS, we applied a numerical spatial low-pass filter on the DM command at the FPM cutoff frequency. The corrections were performed on DM1 as well. As it is currently not possible to run several control loops in parallel on HiCAT, the following mode of operation is used: first, the stroke minimization algorithm provides DM1 and DM2 commands that produce a DH in the source image in the final focal plane; second we calibrate the ZWFS around this non-flat DM configuration; finally we introduce aberrations on DM1 and try to correct for them with the ZWFS measurements using the closed loop with an interaction matrix using 8 Zernike modes numerically low-pass filtered at the FPM cutoff frequency. This procedure does not allow us to work with the best HiCAT contrasts as WFE might drift between the stroke minimization algorithm and the ZWFS control calibration. 

\begin{figure}
    \centering
    \includegraphics[width=\columnwidth]{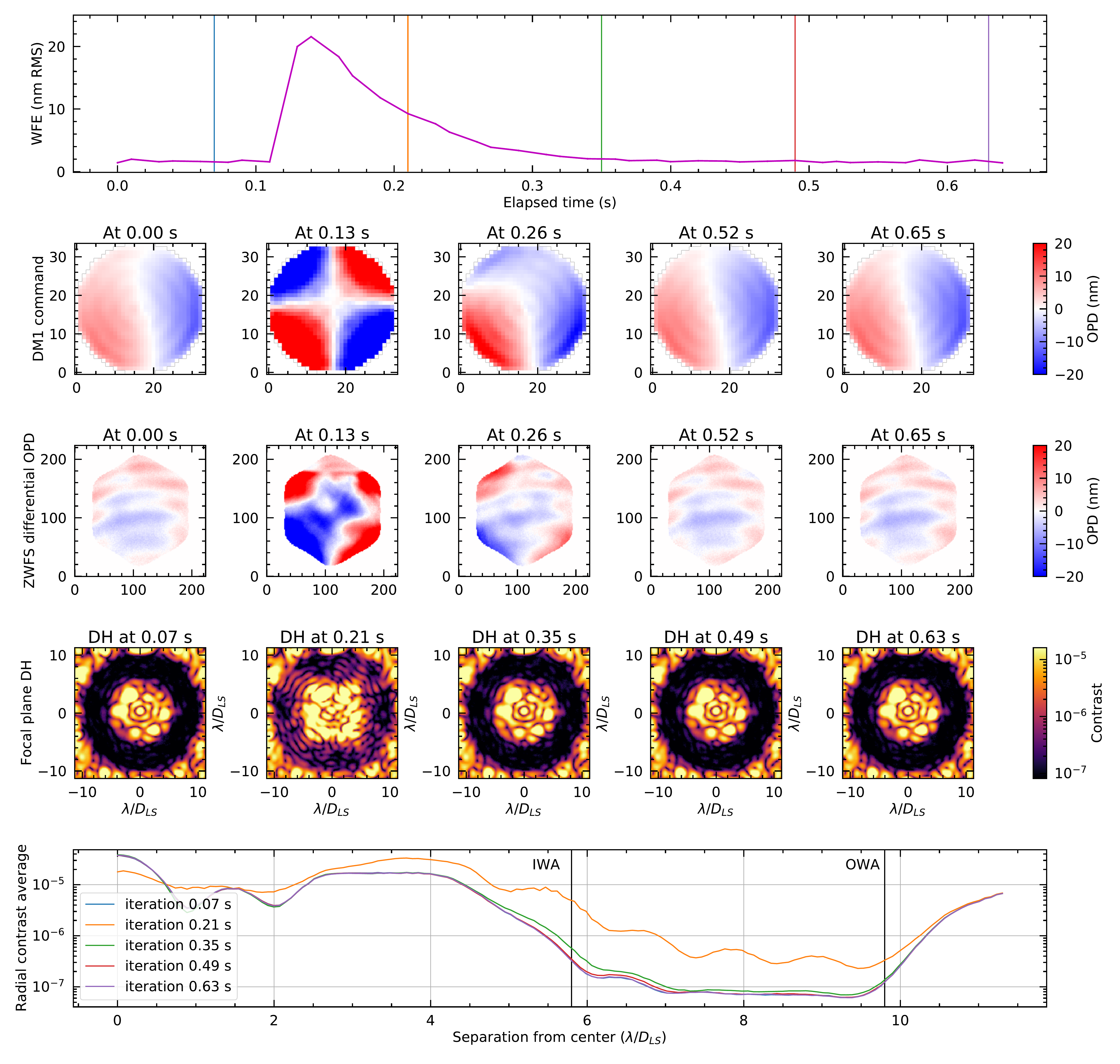}
    \caption{Evolution of HiCAT measurements after introducing a perturbation on DM2 as a function of time. \textit{Top:} Standard deviation in the OPD computed by the ZWFS. \textit{Second row:} Differential commands on DM1 in wavefront error units, the reference DM1 shape for DH being subtracted for readability. \textit{Third row:} OPDs as computed by the ZWFS with analytical reconstruction. \textit{Fourth row:} DH images. \textit{Last row:} Azimutal average of the contrast in fourth row DH images.}
    \label{fig:dh_plot}
\end{figure}

Figure~\ref{fig:dh_plot} sums up this operations. Between 0 s and 0.1 s, the operation is stable, only correcting for turbulence. At 0.12 s, the astigmatism mode of 25 nm RMS is applied on DM1 as visible on the second row. On the third row, the computed OPD shows that part of the defocus is seen by the ZWFS, and is corrected for in a tenth of iterations to come back to a state close to the initial one. In term of intensity in the DH, rows 4 and 5 show an increase of a factor 10 after the introduction of the perturbation. The recovered contrast is identical to the original contrast in the DH that ranges from $5.8\lambda/D_{LS}$ to $9.8\lambda/D_{LS}$, $\lambda$ being the wavelength at 640 nm and $D_{LS}$ the Lyot Stop diameter. To prevent the introduction of aberrations with high spatial frequencies that cannot be measured by the ZWFS the control loop only uses 8 Zernike modes. As visible on the OPDs on the third row, some residual wavefront error in the form of horizontal stripes are not corrected. We attribute this to wavefront error that cannot be projected on the set of modes used. Further work will focus on finding an optimal basis that describes exclusively the space of the filtered aberrations but that does not introduce higher order frequencies unseen by the ZWFS that could degrade the DH.

\section{Conclusion and perspectives}

In this paper we presented some preliminary results of close loop control of the low-order aberrations using the ZWFS as a LOWFS on HiCAT. We show that it is possible to address several questions. First, the correction of low-order modes of the bench internal turbulence, with a reached stability of the order of a few hundreds of picometers. We show that it is possible to use either a Jacobian matrix-based or a phase conjugation phase estimation. Our approach is promising for the wavefront correction and stabilization of HiCAT during other operation, the correction loop being stable down to a few hundreds of picometers over tens of minutes. Finally, this paper presents early results of a closed loop capable of correcting for strong introduced perturbations, here Zernike modes on a DM. Even if the setpoint is the pre-computed DM shapes that produces a DH, the closed loop manages to come back to a state close to the initial state to recover the initial contrast in the DH of the coronagraphic image. 

More parameters will have to be explored to improve these preliminary results. First, the Jacobian matrix approach requires some adjustments: if using Zernike modes as pokes is convenient, we would ideally not introduce frequencies that are not seen by the WFS, here because of spatial filtering. Some trade-offs still have to be found, between the number of controlled modes, gain and type of modes. In term of controller, only an integrator has been considered so far and alternative schemes could improve the stability further: a Proportional Integrator Derivator (PID) or a predictive control algorithm \cite{Gavel2003, Guyon2017}. Furthermore, we assume the pupil intensity is stable over time, which has not been studied in details yet. In addition, we have shown results with HiCAT implementing a CLC coronagraph with a segmented aperture. Finally, the next important step will be to have several loops working in parallel on HiCAT. With the software architecture currently under development to produce and maintain a DH in the coronagraphic image of an observed source, it will be possible in the near future to run simultaneously ZWFS control loop, the tip/tilt control using the quad-cell and the stroke minimization algorithm. The devices coordination would advance the possible simultaneous use of these systems in a space telescope for exo-Earth imaging.

\acknowledgments 
R.P. acknowledges PhD scholarship funding from R\'egion Provence-Alpes-C\^ote d'Azur and Thales Alenia Space. This work was also supported in part by the National Aeronautics and Space Administration under Grant 80NSSC19K0120 issued through the Strategic Astrophysics Technology/Technology Demonstration for Exoplanet Missions Program (SAT-TDEM; PI: R. Soummer)

E.H.P. is supported by the NASA Hubble Fellowship grant \#HST-HF2-51467.001-A awarded by the Space Telescope Science Institute, which is operated by the Association of Universities for Research in Astronomy, Incorporated, under NASA contract NAS5-26555.

\bibliography{report} 
\bibliographystyle{spiebib} 

\end{document}